\documentclass[prb,floatfix,showpacs,superscriptaddress]{revtex4-1}
\usepackage{eurosym}
\usepackage{amsmath}
\usepackage{amssymb}
\usepackage{amsfonts}
\usepackage{graphicx}
\usepackage{dcolumn}
\usepackage{bm}
\usepackage{color}
\usepackage{chngcntr}
\usepackage{changes}
\usepackage{hyperref}
\usepackage{ulem}

\DeclareGraphicsExtensions{.pdf,.png,.jpg,.eps}

\newcommand{\bn}[1]{\mbox{\boldmath $#1$}}

\def\vector#1{\mbox{\boldmath {$#1$}}}

\begin{document}
\title{ Phenomenological model for long wavelength optical modes in transition-metal dichalcogenide monolayer}
\author{C. Trallero-Giner}
\affiliation{Department of Theoretical Physics, Havana University, Havana 10400, Cuba}
\author{E. Men\'{e}ndez-Proupin}
\affiliation{Departamento de F\'{i}sica Aplicada I, Escuela Polit\'{e}cnica Superior, Universidad de Sevilla, Seville, E-41011, Spain}
\affiliation{Departamento de F\'{i}sica, Facultad de Ciencias, Universidad de Chile, Santiago, Chile}
\author{E. Su\'{a}rez Morell}
\affiliation{Departamento de F\'{i}sica, Universidad T\'{e}cnica Federico Santa Mar\'{i}a, Casilla 110-V, Valpara\'{i}so, Chile}
\author{R. P\'{e}rez-\'{A}lvarez}
\affiliation{Universidad Aut\'{o}noma del Estado de Morelos, Ave. Universidad 1001, CP 62209, Cuernavaca, Morelos, M\'{e}xico}
\author{Dar\'{\i}o G. Santiago-P\'{e}rez}
\affiliation{Universidad de Sancti Spiritus ``Jos\'{e} Mart\'{\i} P\'{e}rez'', Ave. de los M\'{a}rtires 360, CP 62100, Sancti Spiritus, Cuba}

\begin{abstract}
Transition metal dichalcogenides (TMDs) are an exciting family of 2D materials; a member of this family, MoS$_2$, became the first measured monolayer semiconductor. In this article, a generalized phenomenological continuum model for the optical vibrations of the monolayer TMDs valid in the long-wavelength limit is developed. Non-polar oscillations involve differential equations for the phonon displacement vector that describe phonon dispersion up to a quadratic approximation. On the other hand, the polar modes satisfy coupled differential equations for the displacement vectors and the inner electric field. The two-dimensional phonon dispersion curves for in-plane and out-of-plane oscillations are thoroughly analyzed. This model provides an efficient approach to obtain the phonon dispersion curves at the $\Gamma$-point of the Brillouin zone of the whole family of TMD monolayers. The model parameters are fitted from density functional perturbation theory calculations. A detailed evaluation of the intravalley Pekar-Fr\"ohlich (P-F) and the $A_1$-homopolar mode deformation potential (Dp) coupling mechanisms is performed. The effects of metal ions and chalcogen atoms on polaron mass and binding energy are studied, considering these two contributions, the short-range Dp and P-F. It is argued that both mechanisms must be considered for a correct analysis of the polaron properties.

\end{abstract}

\date{\today }
\maketitle

\section{\label{Introduction}Introduction}

The emergence of two-dimensional (2D) transition metal dichalcogenide (TMD) materials has encouraged basic research with potentially extraordinary applications in the field of energy storage, biosensors, electronic devices, solar cells, among others (see Refs.~\onlinecite{Berkelbach,Anju,aplications,Kan}). The bulk TMD family of the MX$_2$-type with a transition metal M=Mo, W and a chalcogen X=S, Se, Te consist of two different polytypes with 2H hexagonal honeycomb or rhombohedral symmetries. These structures include semimetals such as MoTe$_2$ and WTe$_2$ or semiconductors like MoS$_2$, MoSe$_2$, WS$_2$, and WSe$_2$.~\cite{Zenga} The 2D molybdenum disulfide MoS$_2$ with semiconducting properties was the first and the most studied single layer member of this family.~\cite{Radi}

Optical phonons of the monolayer (1ML) TMD impact the interband and intraband relaxation processes,~\cite{Song} transport properties,~\cite{Kaasbjerg,transport} photoluminescence,~\cite{IEEE} hot luminescence,~\cite{Urbaszekt} and modulate the  piezoelectricity effect.~\cite{nature} In general in any comprehensive study of dielectric and electronic properties of 2D TMDs the role of the optical phonons needs to be addressed.~\cite{Pike}
In the phonon spectra of 1ML TMDs we can find six optical branches at the $\Gamma$-point of the Brillouin zone (BZ) that are classified in terms of the irreducible representation of the point group $D_{3h}$~\cite{Dresselhaus,XinZhang,7} with symmetries $E^{''}$ (LO$_1$ and TO$_1$), $E^{'}$ (LO$_2$ and TO$_2$), $A_1$ (ZO$_2$), and  $A_2$ (ZO$_1$) which correspond to two in-plane longitudinal (LO$_1$ and LO$_2)$, two in-plane transverse (TO$_1$ and TO$_2$) and two out-of-plane (ZO$_1$ and ZO$_2$) optical vibrations.
 The development of analytical methods together with \emph{ab initio } calculations allow to describe optical oscillations in 2D materials and understand phonon dispersion laws~\cite{Zhang} as well as the electron-phonon interaction.~\cite{PhysRevB.94.085415} A phenomenological continuum model should contain the symmetry properties of the structure and the main characteristics of the optical modes under study.
 In this paper, we present a long wavelength phenomenological continuum approach for the six optical phonon branches of the 2D TMD family.
 The model takes into account the dispersion up to quadratic dependence on the phonon wave vector and their symmetry properties.
Our model assumes in-plane isotropy, as the anisotropy derived from the $D_{3h}$ symmetry is negligible in the dispersion laws for long wavelength.
 This results in six analytical solutions for the optical branches. The parameters of the model were fitted from {\it ab initio} calculations of the phonon spectra. The polaronic corrections are also obtained considering both Fr\"ohlich and short-range deformation potential interactions, emphasizing the important role of the $A_1$-homopolar mode.
 Our model provides an easy approach to get a precise descriptions of the in-plane and out-of-plane phonon properties of all the TMD family.

 The article is organized as follows: In section \ref{Electronphonon} we introduce the phenomenological model starting from a general equation of motion for the relative displacement vector
 of the atoms involved in the phonon oscillations. In section \ref{Abinitio} we describe the procedure we followed with the \textit{ab initio} calculations of the TMD family. In section \ref{fitting} we list all the parameters of our phenomenological model for all the family of TMD obtained from fitting \textit{ab initio} calculations. We elaborate on LO$_2$-TO$_2$ splitting found in the optical modes. In section \ref{Eletronphononinteraction} we derived analytically two main mechanisms responsible for the electron-phonon interaction: Pekar-Fr\"olich and $A_1$-homopolar deformation potential interactions. In section \ref{2DPolaron} employing both Hamiltonians, we explore the polaron properties and obtain the polaron mass and binding energy. Finally, we expose our conclusions in section \ref{Conclusion}.

\section{\label{Electronphonon}Phenomenological model}
Lets recall first some basic concepts of the phenomenological continuum approach adapted to 2D TMD layers.~\cite{PhenoModel}
 We will assume the following general equation of motion for the relative displacement vector, $\bn{u}=(\bn{u}_{\bn{\rho}}(\bn{\rho}),u_z(\bn{\rho}))$, of the atoms involved
valid in the long wavelength limit for the optical phonon branches near the $\Gamma$-point of the BZ,
\begin{eqnarray}
\rho_{m}(\omega^2-\omega_{0}^2) \vector{u} = - \alpha \vector{E}({\bn{\rho},z=0)} \mp \rho_{m}\beta_{1}^2 \nabla(\nabla\cdot\vector{u})
                                            \mp \rho_{m}\beta_{2}^2 \nabla\times\nabla\times\vector{u}\;,
\label{EqMotion2}
\end{eqnarray}
where $\omega_0$ is the natural optical frequency at the in plane phonon wave vector $\bn{q}=\bn{0}$, $\vector{\rho}$ the in-plane radius vector,
$\rho_{m}$ the 2D reduced mass density of the ions involved,  $\bn{E}(\bn{\rho},z=0)$ the 2D macroscopic electric field, $\alpha$ the coupling constant between the mechanical oscillations, $\bn{u}$, and the field $\bn{E}$, $\beta_{1}$  and $\beta_{2}$ are responsible for the quadratic parabolic behavior of the phonon dispersion law. The coefficients $\alpha$, $\beta_{1}$ and $\beta_{2}$ are phenomenological parameters linked to the symmetry of the phonon and obtained by fitting the phonon dispersion law $\omega(\bn{q})$ (experimentally or by \emph{ab initio} calculations). The choice of sign, $+$ or $-$, depends on the curvature of the mode under consideration.

Equation~(\ref{EqMotion2}) is decoupled into two independent equations for in-plane and out-of-plane motions with vector displacement
$\bn{u}_1=(\bn{u}_{\bn{\rho}}(\bn{\rho}),0)$
and $\bn{u}_2=(0,u_z(\bn{\rho}))$, respectively. In consequence, we get four phonon dispersion relations for in-plane longitudinal (LO) and transverse (TO)  branches with symmetry $E^{''}$(LO$_1$,TO$_1$) and $E^{'}$(LO$_2$,TO$_2$) as well as for the two out-of-plane phonon vibrations $A_1$(ZO$_2$) and $A_2$(ZO$_1$).

\subparagraph{\label{ZO1} ZO$_1$-phonon.}
The  optical mode $ A_2$(ZO$_1$) oscillates out-of-plane with a relative vector displacement $\vector{u}_2=\vector{u}_{_{\textrm{ZO}_{1}}}$ parallel to $z$-axis with a reduced mass $\mu^{-1}=m^{-1}_{_\textrm{M}}+(2m_{_\textrm{X}})^{-1}$. Due to the opposite motions of the chalcogens (X$_i$, $i=$1,2) and metal (M) ions the ZO$_1$-mode is polar. Thus, the equation of motion~(\ref{EqMotion2}) is reduced to

\begin{eqnarray}
 \rho_{m}(\omega^2-\omega_{0A_2}^2) \vector{u}_{_{\textrm{ZO}_{1}}} = - \alpha_{_{\textrm{ZO}_{1}}} \vector{E}_z(\vector{\rho},0)
                                            \mp \rho_{m}\beta_{_{\textrm{ZO}_{1}}}^2 \nabla\times\nabla\times\vector{u}_{_{\textrm{ZO}_{1}}}\;,
\label{EqA2}
\end{eqnarray}

\noindent where $\alpha_{_{\textrm{ZO}_1}}$  is the coupling constant between the displacement $\vector{u}_{_{\textrm{ZO}_1}}$  with the $z$-component of the macroscopic electric field, $\vector{E}_z$.
For the solution of Eq.~(\ref{EqA2}) it is necessary to employ the Maxwell equation
\begin{equation}
\nabla\cdot(\vector{E}_z+4\pi \vector{P}_z)=0\;,
\label{Maxwell}
\end{equation}

\noindent where $\vector{P}_z$ is the polarization out-of-plane with the following constitutive relation

\begin{equation}
\vector{P}_z(\vector \rho,z)=[\alpha_{_{\textrm{ZO}_1}}\vector{u}_{_{\textrm{ZO}_1}}(\vector{\rho})+\chi_e \vector{E}_z(\vector \rho,0]p(z)\;,
\label{Polaz}
\end{equation}

\noindent being $\chi_e$ the out-of-plane electronic susceptibility~\cite{Zhang} and $p(z)$ the profile of polarizability density along $z$-direction.~\cite{Kaasbjerg,Rubio,PhysRevB.94.085415}
Here, we consider $p(z)$ to be uniform over the monolayer of thickness $d$, i.e. $p(z)=1/d$ for $|z|\le d/2$ and $0$ otherwise.

The dispersion relation for the out-of-plane ZO$_1$-phonon is obtained by solving the two coupled equations (\ref{EqA2}) and (\ref{Maxwell}) for  $\vector{u}_{_{\textrm{ZO}_1}}$ and $\vector{E}_z$. These  equations represent a generalization of the Born-Huang equations ~\cite{Huang,PhenoModel}
for 2D TMD materials including the quadratic dependence on $\vector{q}$ of the optical mode.  Using $\vector{E}=- \nabla \varphi (\vector{\rho},z)$
with $\varphi (\vector{\rho},z)=\varphi_0 (z)e^{i\small{\vector{q}}\cdot \small{\vector{\rho}}}$
and $\vector{u}_{_{\textrm{ZO}_1}}=u_{0z}e^{i\small{\vector{q}}\cdot \small{\vector{\rho}}}\vector{e}_{z}$, Eq.~(\ref{Maxwell}) is rewritten as

 \begin{eqnarray}
\left[\frac{d^2}{dz^2}-q^2\right]\varphi_0 (z)=4\pi \frac{dp}{dz}\left[ \alpha_{_{\textrm{ZO}_1}}u_{0z}-\chi_e \frac{d\varphi_0 (0)}{dz}\right]\;.
\label {TTTP1}
\end{eqnarray}
 Taking
 \begin{eqnarray}
\varphi_0 (z)=\frac{1}{2\pi}\int_{-\infty}^{\infty}\bar{\varphi}(q_z)e^{iq_z z}dq_z\;,
\label {varphi}
\end{eqnarray}

 \noindent follows
 \begin{eqnarray}
 \bar{\varphi}(q_z)=-\frac{8\pi i}{d}\frac{\sin (q_zd/2)}{q^2+q_z^2}\left[ \alpha_{_{\textrm{ZO}_1}} u_{0z}-\chi_e \frac{d\varphi_0 (0)}{dz}\right ]\:.
\label {varphiq}
\end{eqnarray}

 \noindent Using Eqs.~(\ref{varphi}) and (\ref{varphiq}) we obtain
  \begin{eqnarray}
\frac{d\varphi_0 (0)}{dz}=\frac{4\pi}{d}\left[\alpha_{_{\textrm{ZO}_1}} u_{0z}-\frac{d\varphi_0 (0)}{dz}\chi_e \right ]e^{-\frac{qd}{2}}\:,
\label {varphi0}
\end{eqnarray}
in consequence

 \begin{eqnarray}
\frac{d\varphi_0 (0)}{dz}=\frac{4\pi}{d}
\frac{\alpha_{_{\textrm{ZO}_1}}e^{-\frac{qd}{2}}}{\left[1+\frac{4\pi}{d}\chi_e e^{-\frac{qd}{2}}\right ]}u_{0z} \:,
\label {varphi1}
\end{eqnarray}
and for the electric field
 \begin{eqnarray}
 \vector{E}_z=-\frac{4\pi}{d}
\frac{\alpha_{_{\textrm{ZO}_1}}e^{-\frac{qd}{2}}}{\left[1+ \frac{4\pi}{d}\chi_ee^{-\frac{qd}{2}}\right ]}\vector{u}_{_{\textrm{ZO}_1}}\;.
\label {EZ}
\end{eqnarray}
Inserting Eq.~(\ref{EZ}) into Eq.~(\ref{EqA2}) we have
  \begin{eqnarray}
  \rho_{m}(\omega^2-\omega_{0A_2}^2) \vector{u}_{_{\textrm{ZO}_1} }= \left[\frac{4\pi}{d}\frac{\alpha_{_{\textrm{ZO}_1}}^2e^{-\frac{qd}{2}}}{\left(1+\frac{4\pi}{d}\chi_e e^{-\frac{qd}{2}}\right )}\mp \rho_{m} \beta_{_{\textrm{ZO}_1}}^2 q^2\right]\vector{u}_{_{\textrm{ZO}_1}}\;,
\label {UZO1}
\end{eqnarray}
 hence, the phonon dispersion for the $A_2$-mode is described  by
  \begin{eqnarray}
  \omega^2=\omega_{0A_2}^2 + \frac{4\pi}{\rho_{m}d}
 \frac{\alpha_{_{\textrm{ZO}_1}}^2e^{-\frac{qd}{2}}}{\left(1+\frac{4\pi}{d}\chi_e e^{-\frac{qd}{2}}\right)}\mp \beta_{_{\textrm{ZO}_1}}^2 q^2\;.
\label {UZO11}
\end{eqnarray}

\noindent In the long wavelength $e^{-\frac{qd}{2}} \approx 1$ and $ \omega(q)$ is reduced to

 \begin{eqnarray}
 \omega^2=\omega_{0A_2}^2 + \frac{4\pi}{\rho_{m}d}
 \frac{\alpha_{_{\textrm{ZO}_1}}^2}{\left(1+\frac{4\pi}{d}\chi_e \right)}\mp \beta_{_{\textrm{ZO}_1}}^2 q^2\;.
\label {UZO111}
\end{eqnarray}

\noindent The role of the coupling constant term  $\alpha_{_{\textrm{ZO}_1}}$ and the electronic susceptibility
$\chi_e$ is to renormalize the ZO$_1$ intrinsic oscillatory frequency $\omega_{0A_2}$, thus

 \begin{eqnarray}
 \omega^2=\omega_{A_2}^2 \mp \beta_{_{\textrm{ZO}_1}}^2q^2\;.
\label {omegaAA2}
 \end{eqnarray}
 We take  $\omega_{A_2}^2=\omega_{0A_2}^2 +  4\pi \alpha_{_{\textrm{ZO}_1}}^2/\left[\rho_{m}d (1+\frac{4\pi}{d}\chi_e )\right]$ equal to the frequency value at $q=0$ provided by {\it ab initio } calculations.

\subparagraph{\label{A11} ZO$_2$-phonon.} The out-of-plane nonpolar  $A_1(\textrm{ZO}_2)$ phonon is Raman active with normal vector displacement $\vector{u}_2=\vector{u}_{_{\textrm{ZO}_2}}(\vector{\rho})$ describing the contrary motion of the two chalcogen atoms X$_i$ in the unit cell and $\mu=m_{_{\textrm{X}}}/2$.  The oscillation is independent of the electric field $\vector{E}$ and it is known as homopolar or breathing mode. The mechanical equation of motion for $\vector{u}_{_{\textrm{ZO}_2}}$ is reduced to
\begin{eqnarray}
 (\omega^2-\omega_{A_1}^2) \vector{u}_{_{\textrm{ZO}_2}} =
                                            \mp \beta_{_{\textrm{ZO}_2}}^2 \nabla\times\nabla\times\vector{u}_{_{\textrm{ZO}_2}}\;,
\label{EqA1}
\end{eqnarray}

\noindent with the dispersion relation
\begin{equation}
\omega^2=\omega^2_{A_1} \mp k^2\beta_{_{\textrm{ZO}_2}}^2\;,
\label{Nonpolaz}
\end{equation}
and $\omega_{A_1}$ the frequency at $\bn{q}=\bn{0}$.

\subparagraph{\label{E1mode}Phonon with symmetry $E^{'}$.} Long wavelength longitudinal (LO$_2$) and transverse (TO$_2$) optical phonons with symmetry $E^{'}$ are degenerate at $\Gamma$. The in-plane phonons correspond to the motion of the positive ion relative to the negative ions in the layer and the LO$_2$-mode is due to the longitudinal
displacement with the M-atom in phase opposition to the X$_i$-atoms, $\vector{u}_1=\vector{u}_{E^{'}}$, $\mu^{-1}=m^{-1}_{_\textrm{M}}+(2m_{_{\textrm{X}}})^{-1}$. Due to the translational symmetry on the plane  $\vector{u}_{E^{'}}= \vector{u}_{0E^{'}}e^{i\small{\vector{q}}\cdot\small{\vector{\rho}}}$, $\vector{E}(\vector{\rho},z=0)= \vector{E}(\vector{q})e^{i\small{\vector{q}}\cdot\small{\vector{\rho}}}$, and Eq.~(\ref{EqMotion2}) is re-written as

\begin{eqnarray}
\rho_m( \omega^2-\omega_{E^{'}}^2) \vector{u}_{0E^{'}} =-\alpha \vector{E}(\vector{q})  \pm \rho_m\beta_{_{\textrm{LO}_2}}^2 \vector{q} (\vector{q}\cdot \vector{u}_{0E^{'}})
                                           \pm \rho_m \beta_{_{\textrm{TO}_2}}^2 \left[-\vector{q}(\vector{q}\cdot \vector{u}_{0E^{'}})+q^2\vector{u}_{0E^{'}}\right]\;.
\label{EqE3}
\end{eqnarray}

\noindent In addition, the Eq.~(\ref{EqE3}) is supplemented by the Maxwell  equation $\nabla\cdot \vector{D}=0$, where the induction vector $\vector{D}=\vector{E}(\vector{\rho},z)+4\pi\vector{P}(\vector{\rho},z)$ and for the macroscopic polarization we have the constitutive relation

\begin{equation}
\vector{P}(\vector{\rho},z)=\left[\alpha  \vector{u}_{E^{'}}(\vector{\rho})+\alpha_2 \vector{E}(\vector{\rho},0)\right]p(z)\;,
\label{PolazE1}
\end{equation}
where $\alpha_2$ is the polarizability in the plane.
Using $\vector{E}(\vector{\rho},z)=-\nabla \varphi(\vector{\rho},z)$ follows the Poisson' equation
\begin{eqnarray}
\nabla^2 \varphi+ 4\pi \alpha_2 \nabla^2 _{\vector{\rho}} \varphi _2(\vector{\rho})p(z)= 4\pi \alpha \nabla\cdot[\vector{u}_{E^{'}}(\vector{\rho})p(z)]\;,
\label {TTTP}
\end{eqnarray}
with $\varphi _2(\vector{\rho})=\varphi (\vector{\rho},z=0)$.

For searching the solution of the system of equations (\ref{EqE3}) and (\ref{TTTP}) we take the Fourier transform

\begin{eqnarray}
 \varphi (\vector{\rho},z) = \frac{1}{(2\pi)^3 }\int \bar{\varphi} (\vector{q},q_z)e^{i(\small{\vector{q}.\vector{\rho}}+q_zz)}d^2qdq_z\;.
 \label {FT}
\end{eqnarray}
Inserting (\ref{FT}) into (\ref{TTTP}) we obtain

\begin{eqnarray}
\bar{\varphi}(\vector{q},q_z)&=-&\frac{4\pi}{q_{\perp}^2+q_{z}^2}
\left[\alpha F(\vector{q})
             +\alpha_2 q^2\bar{\varphi}(\vector{q})\right]f\left(\frac{q_{z} d}{2}\right)\;,
\label{SHED5}
\end{eqnarray}
where $f(z)=\sin (z)/z$ and
\begin{eqnarray}
F(\vector{q})=\int \nabla \cdot [\vector{u}_{{E^{'}}}(\vector{\rho})] e^{-i\vector{q}\cdot\vector{\rho}}d^2\rho\:.
\label{FK1}
\end{eqnarray}
\nonumber Integrating Eq.~(\ref{SHED5}) by $dq_z/(2\pi)$ we get that the Fourier transform of the electrostatic potential in the plane $\varphi_2(\vector{\rho})$, takes the form

\begin{eqnarray}
\bar{\varphi}(\vector{q})&=-&\frac{4\pi}{q^2d}
\left[\alpha F(\vector{q})
             +\alpha_2 q^2\bar{\varphi}(\vector{q})\right]\left[1-e^{-qd/2}  \right]\;.
\label{SHED6}
\end{eqnarray}
In the long wave limit $qd/2\ll1$ the function $(1-e^{-qd/2})/(q^2d)$ approaches to $1/(2q)$, hence,~\cite{Note}

\begin{eqnarray}
\bar{\varphi}(\vector{q})&=-&\frac{2\pi}{q}\frac{\alpha F(\vector{q})}{1+2\pi \alpha_2 q}\;.
\label{SHED7}
\end{eqnarray}

\noindent For the solution of Eq.~(\ref{EqE3})  we have to evaluate $\vector{E}(\vector{q},z=0)$. From the Fourier transform of the in-plane electric field $\vector{E}(\vector\rho,z=0)=-\nabla\varphi_2(\vector{\rho})$ we have $\vector{E}(\vector{q})=-i\vector{q}\bar{\varphi}(\vector{q})$ and employing Eq.~(\ref{SHED7}) follows

\begin{eqnarray}
\vector{E}(\vector{q})&=& 2\pi i\frac{\vector{q}}{q}\frac{\alpha F(\vector{q})}{1+2\pi \alpha_2 q}\;.
\label{SHED8}
\end{eqnarray}
Using Eq.~(\ref{FK1}) for  $F(\vector{q})$ and that $\vector{u}_{{E^{'}}}=\vector{u}_{0E^{'}}e^{i\vector{q}\cdot\vector{\rho}}$ we have

\begin{eqnarray}
\vector{E}(\vector{q})= -2\pi \alpha \frac{\vector{q}}{q}\frac{\vector{q}\cdot \vector{u}_{0E^{'}}}{(1+2 \pi\alpha_2q)}\;.
\label{Ek}
\end{eqnarray}
From (\ref{EqE3}) and considering $\vector{u}_{E^{'}}=\vector{u}_{_{\textrm{LO}_2}}+ \vector{u}_{_{\textrm{TO}_2}}$, where $\vector{u}_{_{\textrm{LO}_2}}$($\vector{u}_{_{\textrm{TO}_2}}$) is the longitudinal (transverse) vector displacement with $\vector{q} \times \vector{u}_{_{\textrm{LO}_2}} =\vector{0}$ and $\vector{q}\cdot\vector{u}_{_{\textrm{LO}_2}}\ne 0$ ($ \vector{q}\cdot\vector{u}_{_{\textrm{TO}_2}}= 0$ and $\vector{q}\times \vector{u}_{_{\textrm{TO}_2}} \ne \vector{0}$) we obtain that the phonon dispersion relations for LO$_2$ is
\begin{eqnarray}
 \omega^2= \omega_{E^{'}}^2+\frac{2\pi \alpha^2}{\rho_m} \frac{q}{(1+2 \pi \alpha_2q)} \pm\beta_{_{\textrm{LO}_2}}^2q^2,
 \label{LO2}
\end{eqnarray}

\noindent and for TO$_2$
\begin{eqnarray}
 \omega^2= \omega_{E^{'}}^2 \pm\beta_{_{\textrm{TO}_2}}^2q^2\;.
 \label{TO2}
\end{eqnarray}

\subparagraph{\label{A2}Phonons with symmetry $E^{''}$.}

The modes with symmetry $E^{''}$ are responsible of one-in-plane longitudinal (LO$_1$) and one-in-plane transverse (TO$_1$) optical vibrations.
The normal vector displacement $\vector{u}_1=\vector{u}_{{E^{''}}}(\vector{\rho})$ represents the oscillations of the two  X$_i$ ions,
$\mu=m_{_{\textrm{X}}}/2$ describing the contrary motion of X$_i$-ions of the LO$_1$ and TO$_1$ phonons. In consequence, we are in presence of nonpolar  optical branches (independent of the electric field) with the in-plane LO$_1$ and TO$_1$ doubly degenerate phonons at $\Gamma$.  Thus, Eq.~(\ref{EqMotion2}) can be cast as

\begin{eqnarray}
( \omega^2-\omega_{E^{''}}^2) \vector{u}_{{E''}} =   \mp \beta_{_{\textrm{LO}_1}}^2 \nabla (\nabla \cdot \vector{u}_{{E^{''}}})
                                            \mp \beta_{_{\textrm{TO}_1}}^2 \nabla\times\nabla\times\vector{u}_{{E^{''}}}\;,
\label{EqE2}
\end{eqnarray}
$\omega_{E^{''}}$ being the nature frequency of the in-plane phonons at $\Gamma$.
Taking $\vector{u}_{E^{''}}=\vector{u}_{_{\textrm{LO}_1}}+ \vector{u}_{_{\textrm{TO}1}}$, where $\vector{u}_{_{\textrm{LO}_1}}$($\vector{u}_{_{\textrm{TO}_1}}$) is the independent longitudinal (transverse) normal vector displacement, from Eq.~(\ref{EqE2}) follows the phonon dispersion relations for TO$_1$
\begin{eqnarray}
 \omega^2&=& \omega_{E^{''}}^2\pm \beta_{_{\textrm{TO}_1}}^2q^2,
 \label{EqT1}
\end{eqnarray}
and LO$_1$
\begin{eqnarray}
 \omega^2&=& \omega_{E^{''}}^2\pm \beta_{_{\textrm{LO}_1}}^2q^2.
 \label{EqL1}
\end{eqnarray}

\section{\label{Abinitio} First-principles calculations}

To evaluate of the optical phonon dispersion we performed \emph{ab initio} calculations employing density functional theory perturbation (DFPT)~\cite{baronidfpt} within the Quantum ESPRESSO code.~\cite{espresso} DFPT relies upon density functional theory (DFT) evaluation of the electronic ground state. The DFT calculations were made using the optB86b-vdW exchange correlation functional~\cite{vdW-solids,klimes2010-vdw} to properly consider  the van der Waals interaction. The reciprocal unit cell
 has been sampled using a $\Gamma$-centered $12\times 12\times 1$ $k$-point grid.
The spin-orbit coupling is not included in the calculation. The wave functions of the valence electrons and the electronic density are expanded in plane waves restricted by a kinetic energy cutoff of 65~Ry and 650~Ry, respectively.
The valence electrons are typically at the outer $s$ and $p$ shells of the chalcogen,
and the outer $s$ shell, and the first inner $s,p,d$ shell of the transition metal. The effect of inner electrons
is simulated by means of projector augmented wave (PAW) potentials~\cite{paw1} from the PSLibrary.~\cite{pslibrary} The structures have been optimized  with variable-cell relaxation, until the forces and in-plane stress tensor components were smaller than $10^{-3}$ a.u. and 0.01 kbar, respectively.
The out-of-plane unit cell size is kept constant, as the TMD layer is embedded in vacuum. Vacuum is defined setting a perpendicular
cell size, the $c$ interlayer distance, five times the in-plane lattice parameter.  The Coulomb interaction is truncated in the out-of-plane  direction,~\cite{PhysRevB.94.085415} allowing  total energy, forces and stresses to be computed in a two-dimensional framework.
In our DFPT calculations, the dynamical matrices are computed for a $\Gamma$-centered $6\times 6\times 1$ $q$-point grid in the reciprocal lattice unit cell. The threshold for self-consistency has been set at $10^{-17}$. The dynamical matrices at intermediate $q$-points, are computed by Fourier interpolation, phonon frequencies and eigenvectors are computed by diagonalization.

\begin{table}[b!]
\caption{Employed parameters for the evaluation of the out-plane and in-plane dispersion relations as given by Eqs.~(\ref{omegaAA2}), (\ref{Nonpolaz}), (\ref{LO2}), (\ref{TO2}), (\ref{EqT1}), and (\ref{EqL1}). $a$-optimized lattice constant, $c$-interlayer distance,  $d$-thickness of the monolayer, $r_0$= 2$\pi\alpha_2$-screening parameter, and $\epsilon$-dielectric constant. \label{parameters}}
\centering
\begin{tabular}{lccccccc}
\hline\hline
                           & MoS$_2$  & MoSe$_2$ & MoTe$_2$ & WS$_2$ & WSe$_2$ & WTe$_2$  \\
\hline
   $a$ (\AA)            &   3.1635     & 3.2974       &3.5274  & 3.1627    &3.2954    & 3.5296 \\
  $c$ (\AA)             &  15.9300        & 16.4871  & 17.6370  & 15.8137 &16.4772   &17.6481 \\
  $r_0$(\AA)           &  46.0182           & 53.3517  & 68.6562 & 41.8979  & 48.7041 &64.8081 \\
   $d$ (\AA)             &  5.4817              & 5.9712      & 6.6789   & 5.5093 &  6.0065   & 6.6978  \\
$\epsilon$            &  16.8096          & 17.9009   & 20.5752 & 15.2536 &  16.2581 &19.3865 \\
\hline
\end{tabular}
\end{table}

\begin{figure}[ht]
\begin{center}
\includegraphics[scale=0.6,angle=0]{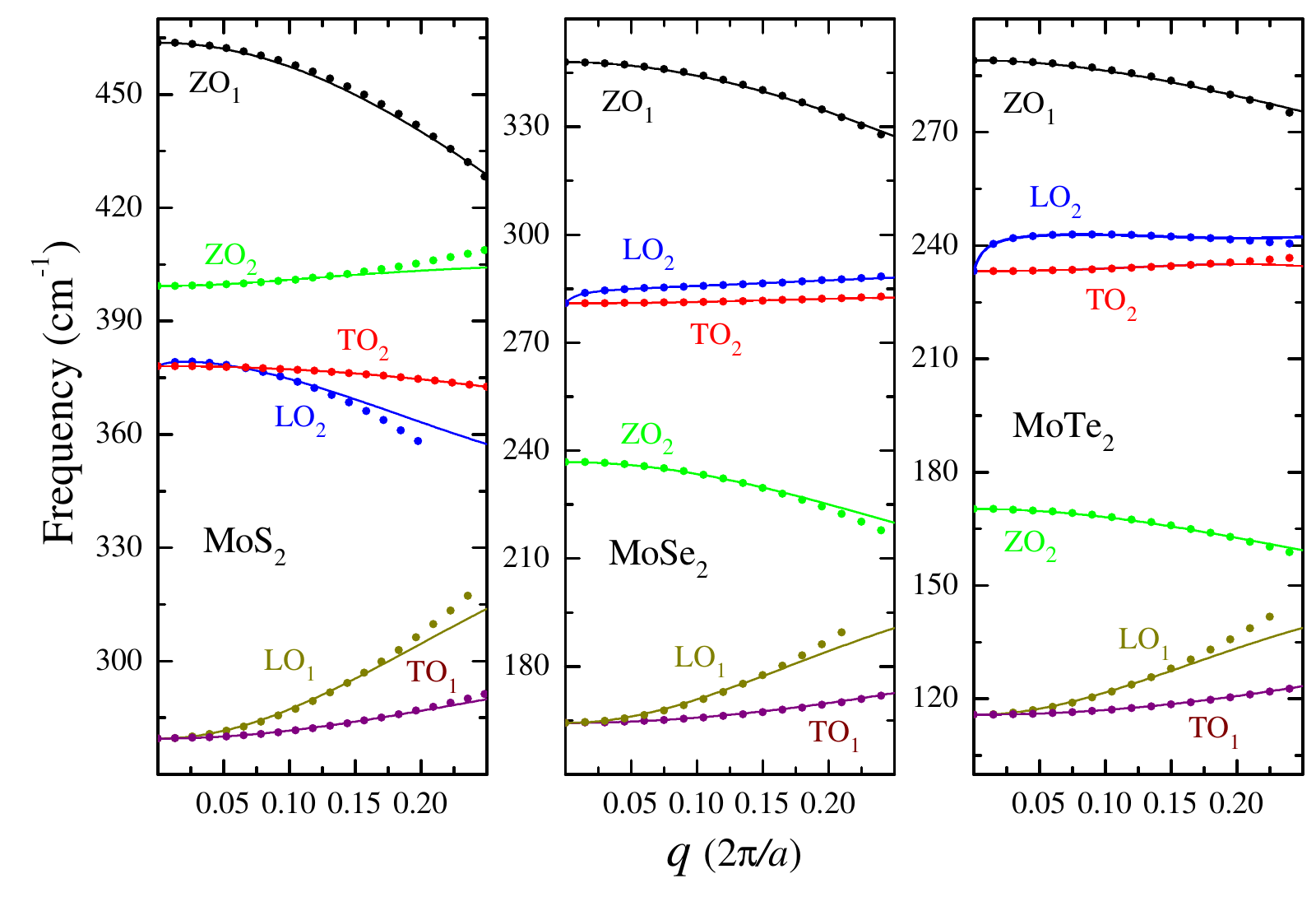}
\end{center}
\caption{\label{FigMoX}  Optical phonon dispersions for 1ML MoX$_2$ (X=S, Se, Te).  Out-of-plane phonons: $A_2$(ZO$_1$) (black) and  $A_1$(ZO$_2$) (green). In-plane-modes with symmetry
$E^{''}$ and $E^{'}$ split into LO$_1$ (dark yellow), TO$_1$ (purple)  and LO$_2$ (blue) and TO$_2$ (red) optical phonons. Straight lines correspond to \emph{ab initio} calculations as explained in Sec.~\ref{Abinitio}, dots represent the dispersion laws obtained by the phenomenological model detailed in Sec.~\ref{Electronphonon}.}
\end{figure}

In Figs.~(\ref{FigMoX}) and~(\ref{FigWX}) we show the optical phonon dispersions, $\omega(q)$, of MoX$_2$ and WX$_2$ (X=S, Se and Te) for phonon wave vectors  $q \le 0.25(2 \pi/a)$. The six optical branches ZO$_1$, ZO$_2$, LO$_2$, TO$_2$, LO$_1$ and TO$_1$ obtained by the \emph{ab initio} procedure along the $\Gamma \to$ K direction of the BZ are shown by straight lines, while the data from the phenomenological model are represented by dots. The parameter values employed for the evaluation of $\omega(q)$ are listed in Table~\ref{parameters}. For the polarizability $\alpha_2$ in Eq.~(\ref{LO2}) we follow the procedure of Ref.~\onlinecite{PhysRevB.94.085415},
where for an isolated TMD the screening parameter $r_0=2\pi \alpha_2$ is reduced to
\begin{eqnarray}
 r_0=\left(\frac{\epsilon}{2}-\frac{1}{3\epsilon}\right)d\;,
 \label{r_o}
 \end{eqnarray}
with the dielectric constant $\epsilon$ evaluated using the standard 3D QE code.~\cite{PhysRevB.94.085415}
In both figures we see the general trend of the phonon frequency with the reduced mass of the atoms involved in vibration, $\omega \thicksim \sqrt{1/\mu}$, i.e., as $\mu$ increases the optical phonon frequency decreases. A noteworthy  change are observed for the out-plane ZO$_2$-phonon with $\mu=m_{_{\textrm{X}}}/2$. Good agreement between the first-principles and phenomenological model calculations are achieved for the phonon dispersions covering approximately 20\% or higher of the BZ near $\Gamma$. In the case of MoS$_2$ the LO$_2$ branch presents an agreement lower than 16\%, while for the WSe$_2$ we achieve a concordance for the ZO$_2$ mode and LO$_2$-phonon smaller than 2\% (for a detailed discussion see below Sec.~{\ref{fitting}).

\begin{figure}[ht]
\begin{center}
\includegraphics[scale=0.6,angle=0]{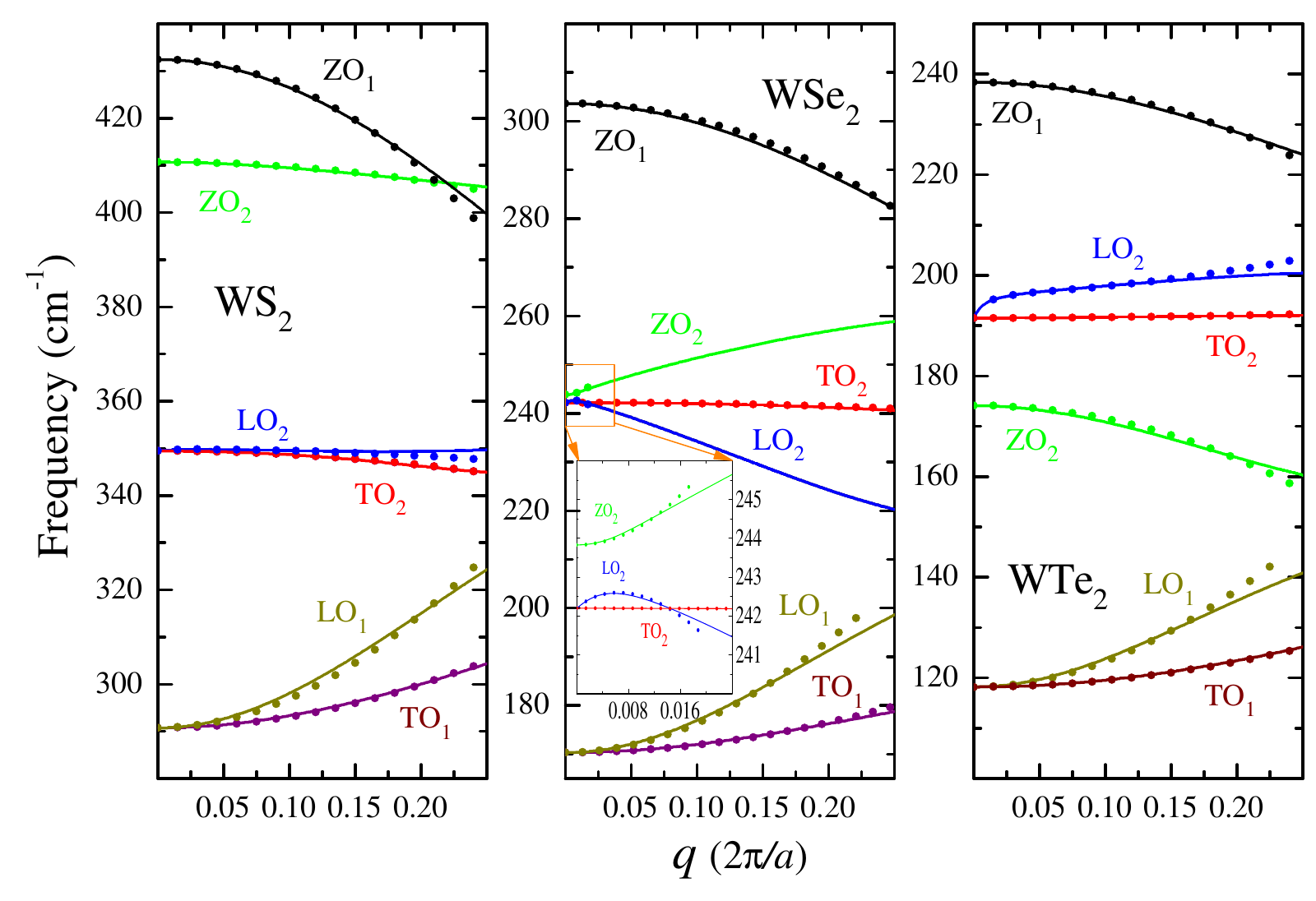}
\end{center}
\caption{\label{FigWX}  The same as Fig.~\ref{FigMoX} for 1ML WX$_2$ (X=S, Se, Te). The inset shows the phonon dispersion of ZO$_2$ , LO$_2$ and TO$_2$ in the small momenta interval for WSe$_2$. }
\end{figure}

\section{\label {fitting} PARAMETERS OF THE MODEL FROM \emph{ab initio} CALCULATIONS}

We are able to extract the characteristic parameters of the phonon frequencies $\omega=\omega(q;\alpha, \beta_{1}, \beta_{2})$ for each phonon symmetry $A_2$, $A_1$, $E^{'}$, and $E^{''}$ at $\Gamma$ by fitting the phonon dispersion laws obtained from
\emph{ab initio}
calculations to the equations proposed in our phenomenological model. These sets
of parameters provide a simple description of the phonon dispersion curves, LO$_2$-TO$_2$ splitting as well as for electron-phonon interaction.
For the fitting procedure we introduce the dimensionless curvature parameters $\mathbb{C}_{\delta}=\pm( 2\pi \beta_{\delta})^2/(a\omega_{\delta})^2$ and  the dimensionless slope $\Xi=(2\pi \alpha)^2/(a \rho_m \omega_{E^{'}}^2)$, where $\delta$=  $A_2$(ZO$_1$), $A_1$(ZO$_2$), $E^{'}$(LO$_2$,TO$_2$), and $E^{''}$(LO$_1$,TO$_1$) takes into account for the phonon symmetry. Employing the analytical phonon dispersions detailed in Eqs.~(\ref{omegaAA2}), (\ref{Nonpolaz}), (\ref{LO2}), (\ref{TO2}), (\ref{EqT1}), and (\ref{EqL1}) those provided by first principle calculations along the $\Gamma \to$ K direction, in Table~\ref{2parameters} we collect the relevant parameters of the six optical phonon branches. For phonon wave vectors $q \le 0.25 (2 \pi/a)$ the two independent directions $\Gamma$ $\to$ K or $\Gamma$ $\to$ M provide practically the same values for both coefficients $\Xi $ and $\mathbb{C}_{\delta}$.

\begin{table}[ht]
\caption{ Phonons data of  MoS$_2$, MoSe$_2$, MoTe$_2$, WS$_2$, WSe$_2$, and WTe$_2$ of 1ML TMDs. The dimensionless curvature $\mathbb{C}_{\delta}=\pm( 2\pi \beta_{\delta})^2/(a\omega_{\delta})^2$ and the dimensionless slope $\Xi=(2\pi \alpha)^2/(a \rho_m \omega_{E^{'}}^2)$ are estimated from the dispersions curves $\omega(q)$ (see Eqs.~(\ref{omegaAA2}), (\ref{Nonpolaz}), (\ref{LO2}), (\ref{TO2}),  (\ref{EqT1}), and (\ref{EqL1}) and the phonon dispersion law obtained by  \emph{ab initio}
calculations for the six optical branches  $\delta$=  $A_2$(ZO$_1$), $A_1$(ZO$_2$), $E^{'}$(LO$_2$,TO$_2$), and $E^{''}$(LO$_1$,TO$_1$). $\omega_{\delta}$ in cm$^{-1}$. \label{2parameters}}
\centering
\begin{tabular}{lccccccc}
\hline\hline
                                     & MoS$_2$    & MoSe$_2$    & MoTe$_2$    & WS$_2$    & WSe$_2$   & WTe$_2$  \\
\hline
$\omega_{{A_2}}$       &  463.73    &  347.97     & 289.03           & 432.45        &303.65         & 238.41\\
$\mathbb{C}_{\textrm{ZO}_1}$   &  -2.38         & -1.95         & -1.62.        & -2.6.              & -2.2            &-2.05\\
\hline
$\omega_{{A_1}}$       &  399.29    &    236.83   & 170.26        & 410.69          &243.83        & 174.12\\
$\mathbb{C}_{\textrm{ZO}_2}$   &  0.78       & -2.68         & -2.25           & -0.48             & 41             & -2.95\\
\hline
$\omega_{{E^{'}}}$       &  378.10    &   280.96    &233.20        & 349.44         & 242.20      & 191.53\\
$\mathbb{C}_{\textrm{LO}_2}$   & -2.85         & 0.54        & -0.60           & -0.22            & -36.5        &  1.1     \\
$\Xi$                              &   1.03      & 3.3            & 12               & 0.23             &  1.2          &  7.0    \\
$\mathbb{C}_{\textrm{TO}_2}$   &   0.46      & 0.23          & 0.52            & -0.43           & -0.16         &  0.13.   \\
\hline
$\omega_{{E^{''}}}$      &  279.51   &  164.34     & 115.80       & 290.73        & 170.31      & 118.18  \\
$\mathbb{C}_{\textrm{LO}_1}$   &   4.7        & 7.45          & 9.84           & 4.3              &   7.2           & 8.8      \\
$\mathbb{C}_{\textrm{TO}_1}$   &   1.39      & 1.61          & 2.12           & 1.6               &   1.84        & 2.17     \\
\hline
\end{tabular}
\end{table}

\subsection{ LO$_2$-TO$_2$ splitting}

An important result valid for resonant Raman scattering,~\cite{Ramanresonant} transport effects,~\cite{transport} polaron effects,~\cite{Ikawa2} magneto-polaron,~\cite{Peeters} exciton-phonon resonances,~\cite{Zimmermann} magneto-Raman scattering~\cite{Iikawa} among other properties, is the splitting between the longitudinal and transverse optical modes.
Using the analytical results for the phonon dispersion law we can study the $\hbar \omega_{_{\textrm{LO}_2}}-\hbar \omega_{_{\textrm{TO}_2}}$ energy splitting and its dependence on the characteristic parameters of the material under consideration.
Figures~\ref{FigLO2TO2MOX2} and \ref{FigLO2TO2WX2} show the dependence of the frequencies $\omega_{_{\textrm{LO}_2}}$ and $\omega_{_{\textrm{TO}_2}}$ on $q$ for the series of MoX$_2$ and WX$_2$ TMD materials.
We observe an excellent agreement of Eqs.~(\ref{LO2}) and~(\ref{TO2}) with our first-principle calculations. In addition, it is clearly seen the linear dispersion relation of the LO$_2$ modes for $1 \gg 2\pi \alpha_2q$ and for the WSe$_2$  the long wave phenomenological model, as dictated by Eq.~(\ref{EqMotion2}), provides good results for the LO$_2$ branch if $q <0.16\times2\pi/a$. From the analytical results displayed in Sec.~\ref{E1mode} we can extract several conclusions. Using Eqs.~(\ref{LO2}) and (\ref{TO2}) the energy of the LO$_2$-TO$_2$ splitting is given by
\begin{eqnarray}
\Delta\omega^{2}_{_{\textrm{LO}_2-\textrm{TO}_2}}=\hbar^2\omega^{2}_{_{\textrm{LO}_2}}-\hbar^2\omega^{2}_{_{\textrm{TO}_2}}=S_{Ph} \frac{q}{(1+2 \pi \alpha_2q)} \pm \hbar^2 \Delta \beta^2 q^2\;,
 \label{omegaLOTO}
 \end{eqnarray}
\noindent with the coefficient $S_{Ph}=2\pi (\hbar \alpha)^2/\rho_m$ and  $\Delta \beta^2=\pm\beta^2_{_{\textrm{LO}_2}} \mp\beta^2_{_{\textrm{TO}_2}}$. Equation~(\ref{omegaLOTO}) shows that for $q \to 0$, $\Delta_{_{\textrm{LO}_2-\textrm{TO}_2}}=\hbar\omega_{_{\textrm{LO}_2}}-\hbar\omega_{_{\textrm{TO}_2}}$ is linear on $q$ with a slope $P=\sqrt{S_{Ph}/2\hbar \omega_{{E^{'}}}}$.
Similar expression has been obtained in Refs.~\onlinecite{Zhang} from a microscopic dipole lattice model  and Ref.~\onlinecite{LO-TO} by first-principles calculations, respectively. In Table~ \ref{SSp} we compare the LO$_2$-TO$_2$ splitting coefficients $S_{Ph}$ obtained from Eq.~(\ref{omegaLOTO}) and $S$ from Ref.~\onlinecite{LO-TO}. Notice that Eq.~(\ref{omegaLOTO}) predicts that  $\Delta_{_{\textrm{LO}_2-\textrm{TO}_2}}$ has a maximum in the neighborhood of $q \sim 0$ for such systems with negative curvature, i.e. $\mathbb{C}_{\textrm{LO}_2}<0$ and $\beta^2_{_{\textrm{LO}_2}}>\beta^2_{_{\textrm{TO}_2}}$ in the phonon dispersion law (MoS$_2$ and WSe$_2$). Hence, the LO$_2$-TO$_2$ splitting may present a maximum at certain $q=q_0>0$ solution of the equation
\begin{figure}[ht]
\begin{center}
\includegraphics[scale=0.6,angle=0]{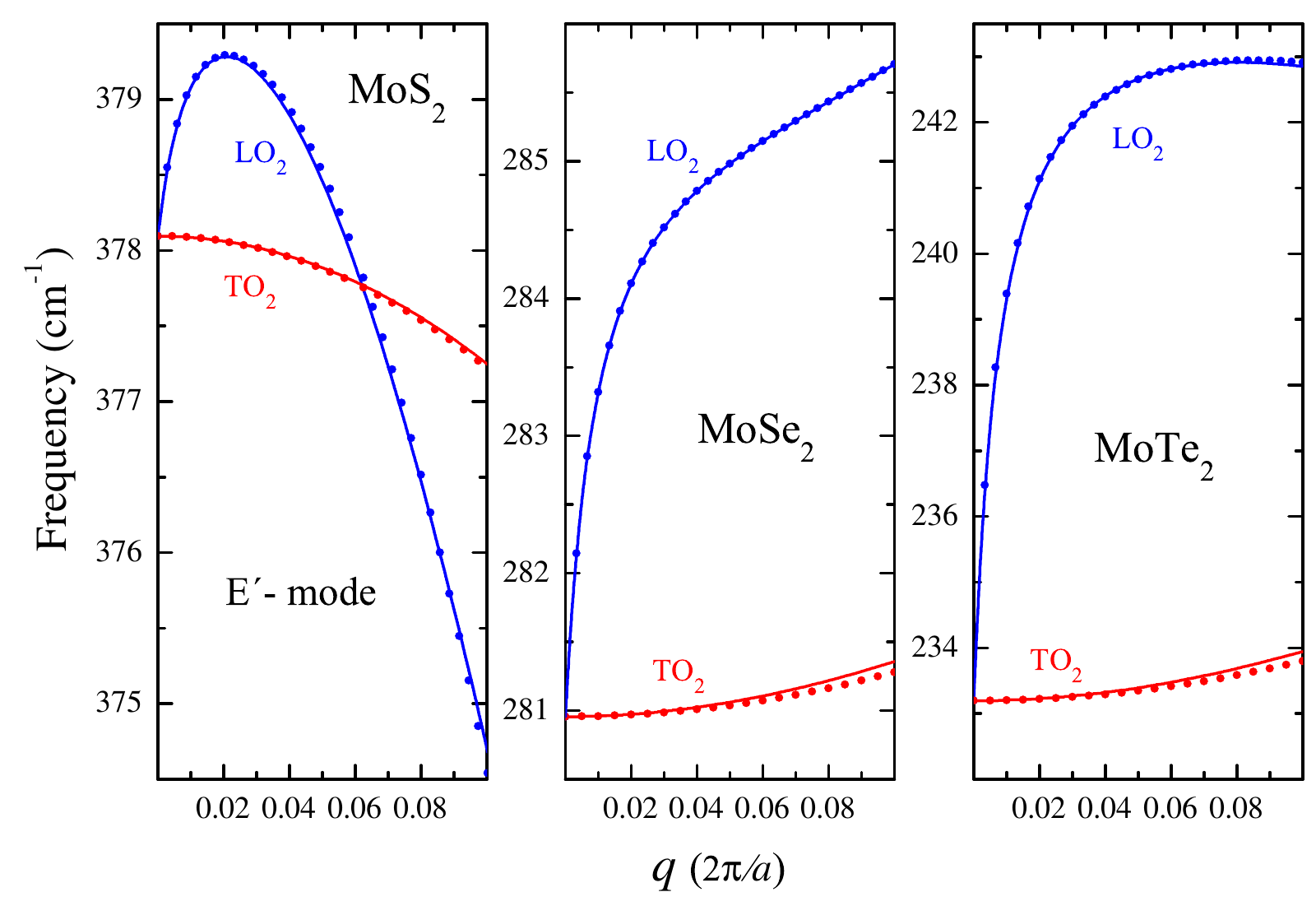}
\end{center}
\caption{\label{FigLO2TO2MOX2}  LO$_2$-TO$_2$ splitting for 1ML MoX$_2$ (X=S, Se, Te). LO$_2$ (blue) and TO$_2$ (red) optical modes. Straight lines correspond to \emph {ab initio} calculations, dots represent the dispersion laws obtained by the Eqs.~(\ref{LO2}) and (\ref{TO2}).
}
\end{figure}
\begin{eqnarray}
q_0(1+ r_{0}q_0)^2=-\frac{S_{Ph}}{2\hbar^2 \Delta \beta^2}\;,
 \label{ko}
 \end{eqnarray}
   \noindent with a maximum  LO$_2$-TO$_2$ splitting $\Delta \omega^2_{Max}=\Delta \omega^2_{_{\textrm{LO}_2-\textrm{TO}_2}}(q=q_0)$.
 As only $q_0>0$ solutions are acceptable, Eq.~(\ref{ko}) does not have solution if $\Delta\beta^2>0$, as in WS$_2$. For the other compounds $\Delta\beta^2<0$ and Eq.~(\ref{ko}) is solved only for MoS$_2$ and WSe$_2$.

 \begin{figure}[ht]
\begin{center}
\includegraphics[scale=0.6,angle=0]{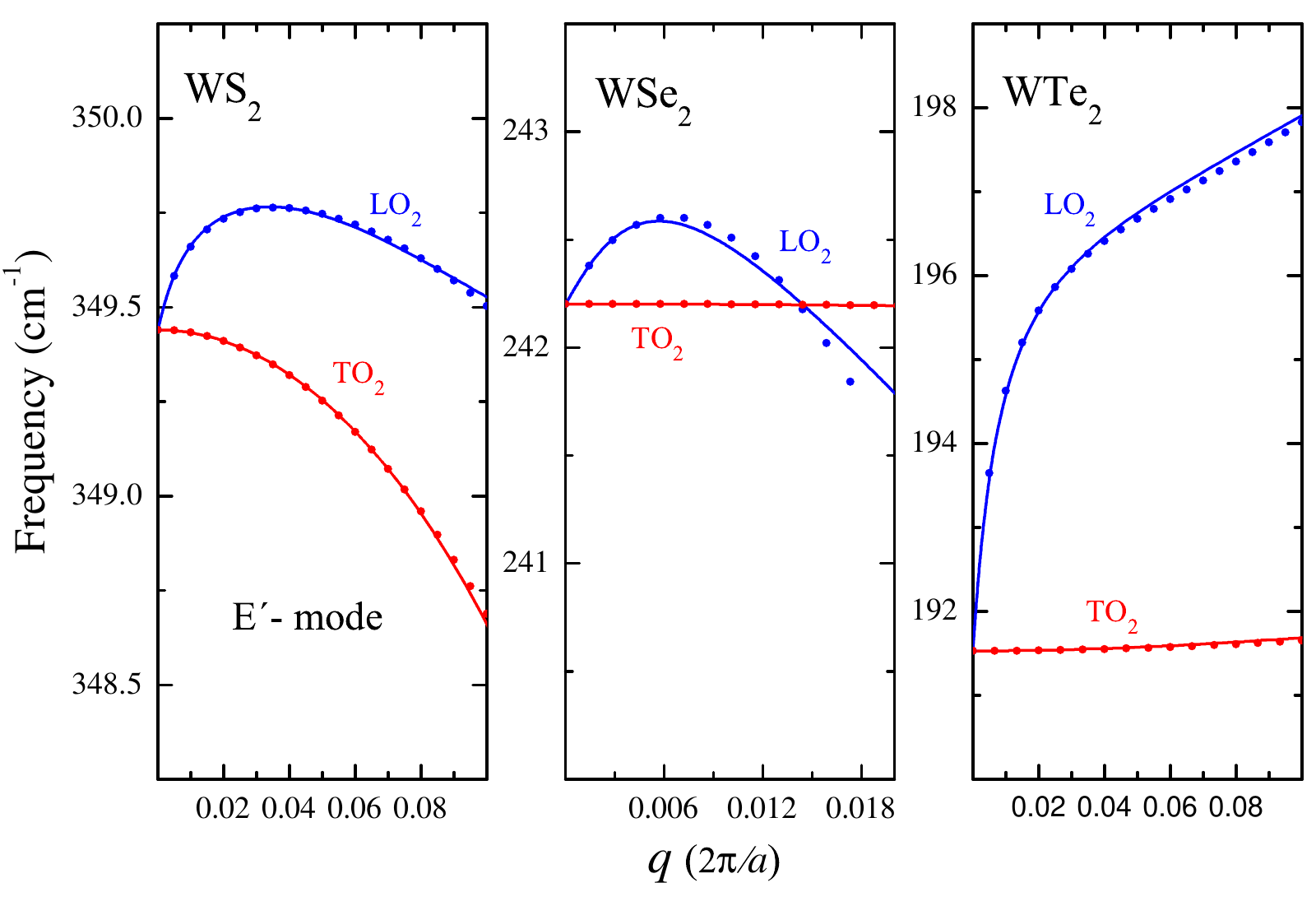}
\end{center}
\caption{\label{FigLO2TO2WX2}  The same as Fig.~\ref{FigLO2TO2MOX2} for WX$_2$ (X=S, Se, Te).
}
 \end{figure}
 \noindent Hence, we learn that as the curvature $\beta_{_{\textrm{LO}_2}}$ increases $q_0 \to 0$. In consequence, the maximum value of the function $\Delta_{_{\textrm{LO}_2-\textrm{TO}_2}}(q_0)$ decreases,  explaining those results for the LO$_2$-TO$_2$ splitting of MoS$_2$ and WSe$_2$ as shown in Figs.~\ref{FigLO2TO2MOX2} and \ref{FigLO2TO2WX2}. Moreover, if the transverse curvature $\mathbb{C}_{\textrm{TO}_2}$  is negative and the TO$_2$-phonon dispersion cannot be considered flat, the LO$_2$-phonon dispersion, Eq.~(\ref{LO2}), has a maximum at $q=q_0$ solution of Eq.~(\ref{ko}). This is the case of the WS$_2$ shown in Fig.~\ref{FigLO2TO2WX2}. We also report the values of  $q_0$ and $\omega_{Max}(q_0)=\omega_{_{\textrm{LO}_2}}(q_0)$ (see Table~\ref{SSp}). It is important to keep in mind that the inclusion of quadratic terms in the equation of motion, proportional to phenomenological parameters $\beta_{_{\textrm{LO}_2}}^2$ and $\beta_{_{\textrm{TO}_2}}^2$, are responsible for the correct description of the $\omega_{_{\textrm{LO}_2}}(q)$ phonon law. From the Table~\ref{SSp} follows that for each series of MX$_2$ (X$=$S, Se, Te), $\Delta \omega^2_{_{\textrm{LO}_2-\textrm{TO}_2}}$ increases as we move from S to Te.
 To explain the dependence of  $\Delta \omega^2_{_{\textrm{LO}_2-\textrm{TO}_2}}$ on chalcogen ions, we realize that the force constant values $\mathbb{F}_{\textrm{X}}$ employed in the first-principle calculations meet the inequality $\mathbb{F}_{\textrm{Te}}<\mathbb{F}_{\textrm{Se}}<\mathbb{F}_{\textrm{S}}$. Therefore, for smaller values of $\mathbb{F}_{\textrm{X}}$ we get larger LO$_2$-TO$_2$ splitting. In addition, if we compare the slopes of the MoX$_2$ set with those of WX$_2$, we have that the Mo compounds have a splitting coefficient $S_{Ph}$ greater than the W counterparts.

 \begin{table}[b!]
\caption{ Comparison of the  LO$_2$-TO$_2$ splitting coefficients $S_{Ph}$ and $S$ reported in Ref.~\onlinecite{LO-TO} for several TDMs. Maximum values of $\omega_{Max}=\omega_{_{\textrm{LO}_2}}(q_0)$ and $q_0$ from Eq.~(\ref{ko}) for MoS$_2$, WS$_2$ and WSe$_2$.\label{SSp}}
\centering
\begin{tabular}{lccccccc}
\hline\hline
                                                               & MoS$_2$   & MoSe$_2$  & MoTe$_2$  & WS$_2$ & WSe$_2$ & WTe$_2$  \\
\hline
$S$ \;\;\; ($10^{-3}$eV$^2$\AA)         &   1.13        & 2.09           & 4.87             & 0.21       &  0.625         & --              \\
\hline
$S_{Ph}$ ($10^{-3}$eV$^2$\AA)              &   1.14       & 2.10           & 5.63              & 0.22       &  0.568        &2.21          \\
\hline
$q_0$\;\; \;\;\;\;\;(\AA $^{-1}$)                                  &    0.043     &   --             &  --                  & 0.07       &  0.011         & --               \\
\hline
$\omega_{Max}$\; (cm$^{-1}$)                 &    379.3      &     --            &  --                 & 349.8     & 242.6          & --               \\
\hline
\end{tabular}
\end{table}

\section{\label{Eletronphononinteraction} Electron-phonon interaction in 2D TMD semiconductors}

The intravalley electron scattering processes in TMD are associated with two main mechanisms: Fr\"ohlich and deformation potential. The first one is linked to internal electrical polarization associated to the optical vibration $\vector{u}$. The LO$_2$-phonon lattice vibration in the TMD layer gives rise to a 2D macroscopic electric field that couples to the charge carriers in the same band. The second is due to the dispersive mechanical nature. The $A_1$-phonon, with the chalcogen  atoms in phase opposition vibrating out-of-plane and fixed Mo or W atoms, is also associated to intraband transitions of carriers. For a study of transport properties~\cite{KIM} or resonant Raman scattering in external fields~\cite{Antagoni}
is mandatory to know the relative contributions of the both interactions.
In this section we present detailed calculations of the Pekar-Fr\"ohlich and deformation potential electron-phonon Hamiltonians.

\subsection{\label{Frohlich1} Pekar-Fr\"ohlich-Type Hamiltonian}

The Pekar-Fr\"ohlich Hamiltonian, $H_F$, is proportional to the LO$_2$-TO$_2$ phonon splitting which is ruled by the coupling constant $\alpha$. This parameter determines the strength of the electron-phonon interaction. For TDM, $H_F$ can be written as

\begin{eqnarray}
H_F &=&-e\varphi_2(\vector{\rho}),
\end{eqnarray}

\noindent where the Fourier transform of the effective potential $\varphi_2$ is

\begin{eqnarray}
\varphi_2 = \frac{1}{(2\pi)^2 }\int \bar{\varphi} (\vector{q})e^{i\small{\vector{q}\cdot \vector{\rho}}}d^2q\;.
\label {2FT}
\end{eqnarray}

\noindent Using Eq.~(\ref{SHED7}) we obtain

\begin{eqnarray}
 \varphi_2 =-\frac{\alpha}{2\pi}\int \frac{F(\vector{q})}{q(1+2\pi \alpha_2 q)}e^{i\small{\vector{q}\cdot\vector{\rho}}}d^2q\;,
 \label {2FT1}
\end{eqnarray}

\noindent with $F(\vector{q})$ is given by Eq.~(\ref{FK1}). The general expression for the LO$_2$-vibrational amplitude can be cast as

\begin{eqnarray}
\vector{u}_{{E^{'}}}= \sum_{\vector{q}}\vector{q}
\left[A_{\vector{q}}e^{i\small{\vector{q}\cdot\vector{\rho}}}+A_{\vector{q}}^*e^{-i\small{\vector{q}\cdot\vector{\rho}}}\right]\;,
\label{U}
\end{eqnarray}

\noindent with the condition $\vector{u}_{{E^{'}}}^*=\vector{u}_{{E^{'}}}$, i.e. $A_{\vector{q}}=-A_{-\vector{q}}^{*}$. The canonical quantum-mechanical version for $\vector{u}_{{E^{'}}}\to \hat{\vector{u}}_{{E^{'}}}$ is obtained by the formal substitution $A_{\vector{q}} \to A_{\vector{q}}\hat{b}_{\vector{q}}$ and $A_{\vector{q}}^* \to A_{\vector{q}}\hat{b}_{\vector{q}}^\dag$,

\begin{eqnarray}
\hat{\vector{u}}_{{E^{'}}}= \sum_{\vector{q}}\vector{q}
A_{\vector{q}}\left[\hat{b}_{\vector{q}}e^{i\small{\vector{q}\cdot\vector{\rho}}}+\hat{b}_{\vector{q}}^\dag e^{-i\small{\vector{q}\cdot\vector{\rho}}}\right]\;,
\label{UO}
\end{eqnarray}

\noindent where $A_{\vector{q}}$ is a real quantity, $\hat{b}_{\vector{q}}$ and $\hat{b}_{\vector{q}}^\dag$, are the second-quantization Bose operator for the annihilation and creation phonon operators, respectively. The operator $\hat{\vector{u}}_{{E^{'}}}(\vector{\rho},t)$ must fulfill the condition for a complete set of functions $[\hat{\Pi}(\vector{\rho},t),\hat{\Pi}(\vector{\rho}^{'},t)]=i\hbar \delta(\vector{\rho}-\vector{\rho}^{'})$ where  $\hat{\Pi}(\vector{\rho},t)=\rho_m \partial \hat{\vector{u}}_{{E^{'}}}(\vector{\rho},t)/\partial t$ is the the momentum density canonically conjugate. Finally

\begin{eqnarray}
\hat{\vector{u}}_{{E^{'}}}= \sum_{\vector{q}}\frac{\vector{q}}{q}\left(\frac{\hbar}{2\rho_m N_cA\omega_{{E^{'}}}}\right)^{1/2}
\left[\hat{b}_{\vector{q}}e^{i\small{\vector{q}\cdot\vector{\rho}}}+\hat{b}_{\vector{q}}^\dag e^{-i\small{\vector{q}\cdot\vector{\rho}}}\right]\;,
\label{UF}
\end{eqnarray}
where $\omega_{{E^{'}}}$ is the phonon frequency as given in Eq.~(\ref{LO2}) and $A$ the area of the unit cell and $N_c$ the number of cell.
Inserting Eq.~(\ref{UF}) into Eqs.~(\ref{FK1}) and Eq.~(\ref{2FT1}) we finally obtain

\begin{eqnarray}
H_F &=&-i\sum_{\vec{k}}\frac{\mathbb{G}_{Ph}}{\sqrt{N{_c}}(1+r_0q)}
\left[\hat{b}_{\vector{q}}e^{i\small{\vector{q}\cdot\vector{\rho}}}+\hat{b}_{\vector{q}}^\dag e^{-i\small{\vector{q}\cdot\vector{\rho}}}\right]\;,
\label{HF1}
\end{eqnarray}
where
\begin{eqnarray}
\mathbb{G}_{Ph}=\left(\frac{2 \pi^2 e^2\hbar \alpha^2}{A\rho_m\omega_{{E^{'}}}}\right)^{1/2}\;,
\label{GPh}
\end{eqnarray}
is the Fr\"ohlich coupling constant obtained by the fitting procedure of the parameter $\alpha$ and the LO$_2$-phonon dispersion.
 In adition, the Fr\"ohlich interaction can be cast in term of the Born effective charges
as given by~\cite{PhysRevB.94.085415,Vogl}
\begin{eqnarray}
\mathbb{G}_{F}=\frac{2\pi e^2}{A}\sum_{i}\frac{\vector{e}_{\vector{q}} \cdot \mathbb{Z}_i \cdot \vector{e}_{\textrm{LO}}}{\sqrt{2M_i\omega_{q}}}\;,
\label{GF}
\end{eqnarray}
where $M_i$ is the mass of atom $i$,  $\mathbb{Z}_i$ the tensors of Born effective charge, $\vector{e}_{\vector{q}}$ the unit vector along the $\vector{q}$-vector, and $\vector{e}_{\textrm{LO}}$ the eigenvector of the LO$_2$-phonon. In Table~\ref{Frohlich} we compare the Fr\"ohlich constants $\mathbb{G}_{Ph}$ and $\mathbb{G}_{F}$ from Eqs.~(\ref{GPh})  and (\ref{GF}), respectively with those values reported in Ref.~\onlinecite{PhysRevB.94.085415}, $\mathbb{G}_{ab}$, for several TMD materials. The small differences between the values of   $\mathbb{G}_{F}$ and $\mathbb{G}_{ab}$ should be awarded to force constants employed in the first principle calculations.

\begin{table}[h]
\caption{ Comparison of the  Fr\"ohlich constants $\mathbb{G}_{Ph}$ from Eq.~(\ref{GPh}) following the fitting procedure, $\mathbb{G}_{F}$ considering our \emph{ab initio} calculations as given by Eq.~(\ref{GF}) and those reported in Ref.~\onlinecite{PhysRevB.94.085415} for several 1ML TDMs . \label{Frohlich}}
\centering
\begin{tabular}{lccccccc}
\hline\hline
Material                       & MoS$_2$    & MoSe$_2$  & MoTe2 & WS$_2$ & WSe$_2$ & WTe$_2$  \\
\hline
$\mathbb{G}_{Ph}$ (eV)   &  0.356    & 0.538        & 0.904  & 0.162   & 0.301  & 0.625 \\
\hline
$\mathbb{G}_{F}$ (eV)     &  0.352    & 0.531        & 0.858  & 0.163   & 0.309  & 0.592 \\
\hline
$\mathbb{G}_{ab}$   (eV)  & 0.334    & 0.502        & 0.819   & 0.140   &0.276  & -----   \\
\hline
\end{tabular}
\end{table}

\subsection{\label{Honopolar} Deformation potential interaction}
Optical homopolar mode changes the electronic energy band due to the mechanical deformation of the atoms in the primitive unit cell. This coupling is known as deformation potential and, in first order of approximation, we will consider this interaction independent of the phonon wave vector. The first-order deformation potential in $\vector{q}$ accounts for electronic intervalley transitions assisted by the ZO$_2$-modes. Assuming the Born-Oppenheimer approximation the electron-phonon contribution to the electronic Hamiltonian of the crystal, $H_c$, is defined as~\cite{Cardona2}
\begin{eqnarray}
H_{Dp} &=& \sum_l \left(\frac{\partial H_c}{\partial \vector{R_l}}\right) \displaystyle \biggl |_0 \cdot \delta \vector{R_l},
\label{HEDP}
\end{eqnarray}
where $\delta \vector{R_l}$ is the atom displacements from the equilibrium. In the long wavelenght limit, the homopolar phonon can be considered as microscopic oscillation within the primitive cell and independent of the phonon wave vector $\vector{q}$ and $\delta \vector{R_l}\approx \vector{u}_{_{\textrm{ZO}_2}}$ the relative vector displacement out-of-plane for the $A_1(\textrm{ZO}_2)$-mode (see Sec~\ref{A11}). Considering a non-degenerate band with energy $E_{\vector{k}}$, the Eq.~(\ref{HEDP}) is reduced to
\begin{eqnarray}
H_{Dp} &=& D_p\vector{e}_z\cdot \vector{u}_{_{\textrm{ZO}_2}}\;,
\label{HEDP2}
\end{eqnarray}
where $D_p$ is the deformation potential characterizing the changes of the electronic energy for $\vector{k}$ near the K-point of the BZ due to the phonon lattice oscillation $ \vector{u}_{_{\textrm{ZO}_2}}$.
Writing $ \vector{u}_{_{\textrm{ZO}_2}}$ in terms of creation and annihilation phonon operator we have
\begin{eqnarray}
H_{Dp}&=&\sum_{\vector{q}}D_p\left(\frac{\hbar}{2\rho_m N_cA\omega_{A_1}}\right)^{1/2}
\left[\hat{b}_{\vector{q}}e^{i\small{\vector{q}\cdot\vector{\rho}}}+\hat{b}_{\vector{q}}^\dag e^{-i\small{\vector{q}\cdot\vector{\rho}}}\right]\;,
\label{UDP}
\end{eqnarray}
with $\omega_{A_1}$ the ZO$_2$-phonon frequency (\ref{Nonpolaz}).

It becomes clear that the electron-phonon coupling in Eq.~(\ref{UDP}) is independent on the phonon wave vector and it corresponds to short-range interaction. In consequences, $H_{Dp}$  is responsable for the intravalley transitions coupling electrons (holes) in the lowest (upper) non-degenerate conduction (valence) band.
The values of the optical-phonon deformation potentials can be extracted by Raman scattering technique,~\cite{Cardona2} transport properties,~\cite{KIM} tight-binding,~\cite{Blacha} or first-principle calculations.~\cite{Kaasbjerg}

\section{\label{2DPolaron}Polaron properties: effective mass and binding energy}

Polaron effects are particularly interesting for cyclotron resonance experiments,~\cite{Devreese} magneto-polaron resonances~\cite{Gurevich, Peeters} and magneto Raman scattering.~\cite{Victor} In 3D semiconductors the binding energy and effective mass depend strongly on the coupling of electrons with LO-phonons at small momenta.  The polaron correction in quasi-2D systems as quantum wells, is a good example as the dimensionality affects the electron-optical phonon coupling.~\cite{Riera}
A straightforward application of the electron-phonon Hamiltonians for 2D TMD developed in Sec.~\ref{Eletronphononinteraction} is the evaluation of the polaron energy of electrons at the K-valley.
Employing the general Green's function formalism, the polaron state is obtained by solving the Dyson equation~\cite{Mahan}

\begin{eqnarray}
G(\vector{k})=G^{(0)}(\vector{k})+G^{(0)}(\vector{k})\sum_{\vector{k}^{'},\vector{q}}S_{\vector{k},\vector{k}^{'}}(\vector{q})G(\vector{k}^{'})\;,
\label{D}
\end{eqnarray}
where G($\vector{k}$) (G$^{(0)}$) is the T$=$0 K one-particle Green$'$s function (unperturbed) for the electron and $S_{\vector{k},\vector{k}^{'}}(\vector{q})$ the irreducible
self-energy. For the unperturbed Green function we know that
\begin{eqnarray}
G^{(0)}=\frac{1}{\varepsilon-\frac{\hbar^2k^2}{2m}+i\delta_0}\;,
\end{eqnarray}
with $m$ the electron effective mass and $\delta_0$ $\to$ 0. To lowest order of electron-optical phonon interaction $S_{\vector{k},\vector{k}^{'}}(\vector{q})=S(\vector{k},\vector{q})\delta_{\vector{k},\vector{k}^{'}+\vector{q}}$, thus, solving Eq.~(\ref{D}) we obtain that
\begin{eqnarray}
G(\vector{k})=\frac{1}{\varepsilon-\frac{\hbar^2k^2}{2m}-S_{\varepsilon}(\vector{k})+i\delta_1}\;,
\label{G01}
\end{eqnarray}
with $S_{\varepsilon}$ the self-energy as given by
\begin{eqnarray}
S_\varepsilon=\sum_{\vector{q}}\frac{|C_{q}|^2}{\frac{\hbar^2}{2m}\left[k^2-|\vector{k}-\vector{q}|^2\right]-\hbar\omega_0+i\delta_1}\;,
\label{S}
\end{eqnarray}
$C_{q}$ being the coupling constant, $\omega_0$ the optical-phonon frequency and the parameter $\delta_1\to 0$.
For the evaluation of the renormalized energy spectrum for conduction electrons, $\varepsilon(\vector{k})$, we take the real part of $G(\vector{k})$.
Considering 3D bulk semiconductors the polaron effective mass and polaron binding energy are $m_{\textrm{3D}}=m/(1-\alpha_{\textrm{3D}}/6)$ and $\Delta \varepsilon=-\alpha_{\textrm{3D}}\hbar \omega_{\textrm{LO}}$, respectively, with $\alpha_{\textrm{3D}}$ the Fr\"ohlich coupling constant and $\omega_{\textrm{LO}}$ the LO phonon frequency.~\cite{Haken}

In 2D TMD the LO$_2$ and $A_1$(ZO$_2$)-modes couple to upper (lower) conduction (valence) band. In consequence, for the evaluation of the self-energy (\ref{S}) we have to add the short-range deformation potential interaction (\ref{UDP}) to the typical Pekar-Fr\"ohlich contribution (\ref{HF1}). It is possible to show that
the polaron effective mass and binding energy are given by  (see Appendix~\ref{appendix}).

\begin{eqnarray}
 m_{p}=\frac{m}{1-\alpha_p}\;,\;\;\;\;\; \Delta \varepsilon_p=-\alpha_F \hbar  \omega_{{E^{'}}}f_1(R_0)-\alpha_{Dp} \hbar \omega_{A_1}\ln2\;,
 \label{MP}
\end{eqnarray}

\noindent with $\alpha_p=2 \alpha_F f_3(R_0)+\alpha_{Dp}/4$. The function $f_i(R_0)$ ($i=1,3$) and the effective coupling constant $\alpha_F$ and $\alpha_{Dp}$ are detailed in the Appendix~\ref{appendix}. The values of $m_p/m_0$ and $\Delta \varepsilon_p$ are summarized in Table~\ref{Pol} for the TMD family, for the evaluation we employ the parameters of Table~\ref{2parameters} and~\ref{Frohlich}.

\begin{table}[h]
\caption{Polaron  mass, $m_p$, binding energy, $\Delta \varepsilon_p$, the Fr\"ohlich 2$\alpha_F f_3(R_0)$, and  deformation potential $\alpha_{Dp}/4$  contributions for  several 1ML TMD. $m$ the electron effective mass at K-point, $m_0$ the bare electron mass and D$_c$ the deformation potential.  \label{Pol}}
\centering
\begin{tabular}{lccccccc}
\hline\hline
Material                       & MoS$_2$         & MoSe$_2$       & MoTe2                                       & WS$_2$                 & WSe$_2$                \\
\hline
$m/m_0$                     & 0.51$^{a}$       &0.64$^{a}$  & 0.56$^{b}$    &  0.31$^{a}$ &0.39$^{a}$       \\
\hline
$D_p$  (eV/\AA)          & 5.8$^{a}$   &  5.2$^{a}$ &  1.34$^{d}$                 &  3.1$^{a}$       & 2.3~$^{a}$            \\
\hline
2$ \alpha_F f_3(R_0)$ & 0.009         &  0.032           & 0.094    &  0.002   &  0.013         \\
\hline
$\alpha_{Dp}/4$           &0.017          &   0.035           &  0.004    &   0.002     & 0.003     \\
\hline
$m_p/m_0$                 &  1.027         &  1.071             & 1.108     &   1.004  &  1.016             \\
\hline
$\Delta \varepsilon_p$ (eV) &  -0.0024 &  -0.0028      & -0.0002  &  -0.0003 &  -0.0003     \\
\hline
$^{a}$ Ref.~\onlinecite{KIM}\;; $^{b}$ Ref.~\onlinecite{PhysRevB.94.085415}\;; $^{d}$ Ref.~{\onlinecite{2Huang}}.\\

\end{tabular}
\end{table}
From the results of Table\ref{Pol} we have that for a given serie of MX$_2$ compounds as the mass of the chalcogen atom X increases, the $m_p$ increases following the same trend of the Fr\"ohlich coupling constant (\ref{GPh}). In addition, it can be seen that the deformation potential plays a crucial role for the binding energy and for the polaron mass in particular for the MoS$_2$ and MoSe$_2$ materials. For the WX$_2$ compounds the deformation potential contribution is weaker than MoX$_2$ ones.

\section{\label{Conclusion} Conclusions}
We implement a continuum phenomenological approach valid for any monolayer TMDs family. The model accurately describes the dispersive phonon spectra of the non-polar and polar optical phonon modes with in-plane (LO$_1$, TO$_1$, LO$_2$, and TO$_2$) or out-plane oscillations (ZO$_1$ and ZO$_2$). We employ \emph{ab initio} calculations to evaluate the characteristic parameters of the phonon dispersion laws up to parabolic term with  the wave vector $\vector{q}$ (see Table~\ref{2parameters}). These results allow to establish the validity framework of our  phenomenological model (see Figs.~\ref{FigMoX}-\ref{FigLO2TO2WX2}).
Under the condition $\beta_{_{\textrm{LO}_2}}>\beta_{_{\textrm{TO}_2}}$ we predict that the LO$_2$-TO$_2$ splitting has a maximum for TMDs with negative curvature in the phonon dispersion. For the case of LO$_2$ and ZO$_2$ oscillations in WSe$_2$, the range of validity of their respective phonon dispersions is restricted to a small interval of $\vector{q}$ ($q<$ 0.016 2$\pi/a$ in Fig.~\ref{FigWX}). A plausible explanation is that our model in (\ref{EqMotion2}}) does not take into account the interplay between branches with different symmetries. The frequency separation between $\omega_{A_1}$ and $\omega_{E^{'}}$ is less than 2 cm$^{-1}$. The results provide us with a complete description of the 2D long-range (\ref{HF1}) and short-range homopolar deformation potential (\ref{UDP}) electron-phonon interactions. We report in Table~\ref{Frohlich} the strength values of the coupling constant needed to obtain analytically the 2D intravalley Pekar-Fr\"ohlich Hamiltonian for the TMDs family.
 Finally, using the electron-phonon Hamiltonians~(\ref{HF1}) and~(\ref{UDP}) we report the polaron properties for TMDs MoS$_2$, MoSe$_2,$ MoTe$_2$, WS$_2$, and WSe$_2$. Consequently, the results detailed in Table~\ref{Pol} show that the contribution of the deformation-potential interaction associated with $A_1$-mode cannot be disregarded. This conclusion can be extended to those processes assisted by intravalley transitions.

\appendix

\section{\label{appendix} Polaron energy}
The total self-energy
$S_{\varepsilon}=S_{\varepsilon}^{(F)}+S_{\varepsilon}^{(Dp)}$
where $S_{\varepsilon}^{(F)}$ corresponds to the Fr\"ohlich interaction while $S_{\varepsilon}^{(Dp)}$ to the mechanical lattice distortion.
\subparagraph{Fr\"ohlich contribution.}
From the Eq.~(\ref{HF1}) follows that $C_{q}=\mathbb{G}_{Ph}/\sqrt{N{_c}}(1+r_0q)$. Thus, the self-energy $S_{\varepsilon}^{(F)}$ can be written as
\begin{eqnarray}
S_{\varepsilon}^{(F)}=-\alpha_F \hbar \omega_{{E^{'}}}F(\kappa,R_o)\;,
\label{SEF}
\end{eqnarray}

\noindent where $\kappa=k/\sqrt{\hbar/2m\omega_{{E^{'}}}}$, $R_0=r_0\sqrt{2m\omega_{{E^{'}}}/\hbar}$,
\begin{eqnarray}
\alpha_F=\frac{\sqrt{3}}{2\pi}\frac{m}{\hbar^2a^{-2}}\frac{\mathbb{G}_{Ph}^2}{\hbar \omega_{{E^{'}}} }\;,
\label{alpha}
\end{eqnarray}
and
\begin{eqnarray}
F(\kappa,R_0)=\int_0^\infty \frac{z}{(1+R_0z)^2\sqrt{(1+z^2)^2-4\kappa^2 z^2}}dz\;.
\label{F}
\end{eqnarray}

\noindent If $\hbar^2k^2/(2m)<\hbar \omega_{{E^{'}}} $ we can approximate the function

\begin{eqnarray}
(1+z^2-4\kappa^2 z^2)^{-1/2} \approx \frac{1}{z^2+1}+\frac{2z^2}{(z^2+1)^3}\kappa^2\;,
\label{AP}
\end{eqnarray}

\noindent and $S_{\varepsilon}^{(F)}$ can be cast as
\begin{eqnarray}
S_{\varepsilon}^{(F)}=-\alpha_F \hbar \omega_{{E^{'}}} [f_1(R_0)+2\kappa ^2f_3(R_0)]\;,
\label{SE2}
\end{eqnarray}

\noindent with
\begin{eqnarray}
f_p=\int_0^\infty \frac{z^p}{(1+R_0z)^2(1+z^2)^p}dz\;.
\label{Ip}
\end{eqnarray}

\subparagraph{Deformation potential.}
For the $A_1$ modes the electron-deformation potential coupling constant is
\
\begin{eqnarray}
C_q^2=\frac{\hbar}{2\rho_m N_cA\omega_q}D_{p}^{2}
\label{Cdp}
\end{eqnarray}
and for the $S_{\varepsilon}^{(Dp)}$ we obtain

\begin{eqnarray}
S_{\varepsilon}^{(Dp)}=-\alpha_{Dp} \hbar \omega_{{A_1}}\left[ \ln 2 +\frac{\kappa ^2}{4}\right]\;,
\label{SE2DP}
\end{eqnarray}
where
\begin{eqnarray}
\alpha_{Dp}=\frac{m}{4\pi\rho_m }\left(\frac{D_p}{\hbar  \omega_{{A_1}}}\right)^2\;.
\label{SEFF}
\end{eqnarray}

\noindent Taking the real part of the Green function (\ref{G01}) we obtain the 2D polaron energy as
\begin{eqnarray}
\varepsilon_{p}(\vector{k})=\frac{\hbar^2k^2}{2m}\left[1-2 \alpha_F f_3(R_0)+\frac{\alpha_{Dp}}{4}\right]-\alpha_F \hbar  \omega_{{E^{'}}}f_1(R_0)-\alpha_{Dp} \hbar \omega_{A_1}\ln2\;.
\label{E2D}
\end{eqnarray}

\acknowledgments
R.P-A thanks CONACyT/Mexico for a sabbatical grant. E.S.M and E.M-P acknowledge support by FONDECYT Chile grant 1170921.


\begin{thebibliography}{1}
\bibitem{Note} This result corresponds to the non-homogeneous solution of Poisson' equation in the long wave limit and considering the TMD isolated. In order to account the presence of the environment, we have to include in Eq.~(\ref{TTTP}) the  spatial dependence of the dielectric constant on $z$, $\varepsilon=\varepsilon_T(z)$, i.e.  $\varepsilon_T=\varepsilon_1$ if $z<-d/2$, $\varepsilon_T=\varepsilon_2$ for $z>d/2$ and equal to $\varepsilon_0$ in the interval $-d/2<z<d/2$. The general solution requires to fulfill the matching boundary conditions at $z=\pm d/2$.~\cite{PhysRevB.94.085415}
\end{thebibliography}
\end{document}